\documentclass[final]{aipproc}

\layoutstyle{6x9}

\usepackage{bm}
\usepackage{amsmath}

\newcommand{\energy}{m_T}

\begin{document}

\title{
Fast Vacuum Decay into Quark Pairs in Strong Color Electric and Magnetic
Fields
}


\author{Y.~Hidaka}{
  address={Department of Physics, Kyoto University, Kitashirakawaoiwake, Sakyo, Kyoto 606-8502, Japan}
,email={hidaka@ruby.scphys.kyoto-u.ac.jp}
}

\author{T. Iritani}{
  address={Department of Physics, Kyoto University, Kitashirakawaoiwake, Sakyo, Kyoto 606-8502, Japan}
  ,email={iritani@ruby.scphys.kyoto-u.ac.jp}
}

\author{H. Suganuma}{
  address={Department of Physics, Kyoto University, Kitashirakawaoiwake, Sakyo, Kyoto 606-8502, Japan}
  ,email={suganuma@ruby.scphys.kyoto-u.ac.jp}
}

\begin{abstract}
We study quark-pair creations in strong color electromagnetic fields. 
We point out that, for massless quarks, 
the vacuum persistency probability per unit space-time volume is 
zero, i.e., the quark-pair creation rate $w$ is infinite, 
in general homogeneous color electromagnetic fields, 
while it is finite when the color magnetic field is absent. 
We find that the contribution from the lowest Landau level (LLL) dominates 
this phenomenon. 
With an effective theory of the LLL projection, 
we also discuss dynamics of the vacuum decay, 
taking into account the back reaction of pair creations.
\end{abstract}

\maketitle


\section{Introduction}
Dynamics in strong (color) electromagnetic fields 
has been an interesting subject in theoretical physics 
\cite{Heisenberg, Schwinger, Schwinger3, Suganuma, Tanji, Iwazaki, Hidaka}. 
Recently, this subject is being paid attention also 
in high-energy heavy-ion collision experiments.
At the so-called Glasma stage~\cite{Glasma} just after the collision,
longitudinal color-electric and color-magnetic fields are expected to be 
produced in the context of the color glass condensate 
of order $1$--$2$GeV in RHIC and $5$GeV in LHC.
In the peripheral collision, a strong magnetic field of order $100$ MeV 
would be also induced. 
Such strong fields will play important role of particle productions 
and thermalization to the quark gluon plasma. 
In this paper, we concentrate on how the strong fields decay into particles.
For this purpose, we consider the Schwinger mechanism for quarks 
in the coexistence of color-electric and color-magnetic fields. 
We will point out that, in the case of massless quarks, 
the vacuum immediately decays 
in general homogeneous (color) electromagnetic fields.

\section{Vacuum decay in color electromagnetic field}
To simplify the situation of particle creations, we consider 
{\it covariantly constant} color electromagnetic fields~\cite{Suganuma} of 
$[D_\mu, F_{\alpha \beta}]=0$, i.e.,
$[D_\mu,\bm{E}]=[D_\mu,\bm{B}]=\bm{0}$, 
where $D_\mu \equiv \partial_\mu-ig A_\mu$ is the covariant derivative 
with the gauge field $A_\mu$. 
We here set $g>0$.
The color electric and magnetic fields are defined as
$\bm{E}^i=F^{i0}$ and $\bm{B}^i=-\epsilon^{ijk}F_{jk}/2$ 
with $F_{\mu\nu}\equiv i[D_{\mu},D_{\nu}]/g$.
This is a gauge-covariant generalization of constant fields in QED, 
$\partial_{\mu} \bm{E}=\partial_{\mu} \bm{B}=\bm{0}$, 
to the non-Abelian fields, and the translational invariance 
leads to this condition.
For the covariantly constant fields, all the components of 
$\bm{E}$ and $\bm{B}$ can be diagonalized to be constant matrices 
in color space by a gauge transformation.
Without loss of generality, 
one can set $\bm{E}=(0,0,E)$ and $\bm{B}=(0,0,B)$ 
in a suitable Lorentz frame.
Here, $B$ and $E$ can be written as Lorentz scalar defined by 
$B, E \equiv [({\cal F}^2+{\cal G}^2)^{1/2} \pm {\cal F}]^{1/2}$ 
\cite{Suganuma} 
with ${\cal F} \equiv F_{\mu\nu}F^{\mu\nu}/4=(\bm{B}^2-\bm{E}^2)/2$ 
and ${\cal G} \equiv F_{\mu\nu}\tilde F^{\mu\nu}/4=\bm{B}\cdot\bm{E}$. 

Like the vacuum decay in QED~\cite{Heisenberg,Schwinger},
we consider the vacuum persistency probability,
\begin{equation}
|\langle\Omega_\mathrm{out}|\Omega_\mathrm{in}\rangle|^2=\exp(-VTw) ,
\end{equation}
where $V$ and $T$ are infinite space volume and time length.
$|\Omega_\mathrm{in}\rangle$  and $|\Omega_\mathrm{out}\rangle$ are
the in-vacuum and the out-vacuum defined 
at $t=-T/2$ and $t=T/2$, respectively.  
$w$ is called ``pair-creation rate'', and 
denotes magnitude of the vacuum decay per unit space-time volume.
The analytic formula of $w$ for the quark-pair creation 
with the quark mass $m$ 
in the covariantly constant color fields is obtained as \cite{Suganuma,Hidaka}
\begin{equation}
\begin{split}
 w
 &=\sum_{N=1}^\infty \mathrm{tr} \left\{\frac{g^2 EB}{4\pi^2}
\frac{1}{N}\coth(\pi N BE^{-1}) e^{-N\pi m^2/gE} \right\} \\
  &=\sum_{s_z=\pm 1/2}\sum_{n=0}^\infty \mathrm{tr}
\left\{ \,\frac{g^2 EB}{4\pi^2}
  \ln\frac{1}{1-\exp{(-\pi \energy^2(n,s_z)/gE})}\right\},
  \end{split}
\label{eq:wqcd}
\end{equation}
where $\energy(n,s_z)\equiv \{2gB(n +1/2\mp s_z)+m^2\}^{1/2}$ 
is the transverse energy, $n$ labels the Landau level,
and $s_z=\pm1/2$ denotes the spin component. 
The trace is taken over the indices of color and flavor.
This is the non-Abelian extension of the formula for QED.
(On the quark mass $m$, chiral symmetry is restored by 
strong (color) electric field \cite{Suganuma}.)
The lowest Landau level (LLL) with $n=0$ and $s_z=1/2$ gives 
$\energy^2(n,s_z)=m^2$, which is independent of $B$, while 
$\energy^2(n,s_z)$ of higher levels includes $O(gB)$ term.
Hence, the LLL state gives the dominant contribution for $gB\gg m^2$.
The LLL contribution for small $m$ is 
\begin{equation}
  w=\mathrm{tr}\left\{ \, \frac{g^2 EB}{4\pi^2}
  \ln \frac{1}{1-\exp(-\pi m^2/gE)} \right\}
  \simeq \mathrm{tr} 
  \left\{\,\frac{g^2EB}{4\pi^2}\ln\frac{gE}{\pi m^2} \right\}.
\end{equation}
As $m\to0$,  $w$ diverges as $w \propto -\ln m\to \infty$.
Note that $w$ diverge, only when the (color) magnetic field coexists, and
the Landau quantization causes the divergence of $w$~\cite{Hidaka}.

\section{Effective theory in a strong magnetic field}
When the (color) magnetic field is strong enough, 
the LLL dominates and hence the LLL projected theory is applicable, 
because of large $m_T(n,s_z)$ for higher Landau levels.
As a remarkable merit of the LLL projected theory, 
we can take into account the back reaction of created particle pairs. 
In this section, we investigate the LLL projected theory 
in QED or Abelianized QCD, 
since the fermion-pair creation formalism in the covariantly constant fields 
in QCD has Abelian nature and is similar to the Schwinger mechanism in QED.
The wave function of the fermion is projected to the LLL state, and 
the LLL wave function in a symmetric gauge of ${\bf A} =(-yB/2,xB/2,0)$ is
\begin{equation}
\phi_l(x,y)=\sqrt{\frac{eB}{2\pi l!}}
\left(\frac{eB}{2}\right)^{\frac{l}{2}}(x+iy)^l
\exp\left(-\frac{eB}{4}(x^2+y^2)\right),
\end{equation}
where $l (\geq 0)$ denotes the angular momentum in $z$ direction for the LLL, 
and the energy is degenerate on $l$.
The projected wave function can be written as
\begin{equation}
\psi(x,y,z,t)= 
\begin{pmatrix}
\sum_l \phi_l(x,y)\varphi_l(t,z)\\
0 
\end{pmatrix} ,
\end{equation}
in a suitable representation,
where $\varphi_l(t,z)$ is the two component Dirac field in 1+1 dimensions.
Then the fermion action of QED in $3+1$ dimensions reduces to 
that of non-Abelian gauge theory in $1+1$ dimensions:
\begin{equation}
S=\int d^4x \bar{\psi}(x)i\gamma^\mu D_\mu\psi(x)
\simeq
\sum_{l,l'}\int dtdz \bar{\varphi}_{l'}(t,z)i\tilde{\gamma}^\mu \tilde{D}^{l'l}_\mu\varphi_{l}(t,z),
\label{eq:action}
\end{equation}
where $\tilde{\gamma}^{~t}$ and $\tilde{\gamma}^{~z}$ are the gamma matrices 
in $1+1$ dimensions and $\tilde{\gamma}^{~x}=\tilde{\gamma}^{~y}=0$.
The covariant derivative is defined by $\tilde{D}^{l'l}_\mu
=\delta^{l'l}\partial_\mu-ie \tilde{A}^{l'l}_\mu$ with
\begin{equation}
\tilde{A}^{l'l}_\mu(t,z)=\int dxdy\phi_{l'}^*(x,y)\phi_l(x,y) A_\mu(x,y,z,t).
\label{eq:gaugeField}
\end{equation}
$\tilde{A}^{l'l}_\mu(t,z)$ turns out the gauge field 
in $U({\cal N})$ gauge theory with 
${\cal N} \equiv eBV_\perp/(2\pi)(=+\infty)$, 
since $\tilde{A}^{l'l}_\mu(t,z)$ is an Hermite matrix, 
$\tilde{A}^{*l'l}_\mu(t,z)=\tilde{A}^{ll'}_\mu(t,z)$, 
and the indices $l$ and $l'$ run from $0$ to ${\cal N}-1$. $V_\perp$ 
is the volume of the transverse directions.
While the original gauge field lives in 3+1 dimensions 
as Abelian gauge field, 
the gauge field in the effective fermion action 
is reduced to non-Abelian gauge field in 1+1 dimensions.
The gauge field can be decomposed to $U(1)$ and $SU({\cal N})$ parts,
which are $\tilde{A}_\mu(t,z) \equiv 
\sum_l\tilde{A}^{ll}_\mu(t,z)/{\cal N}$, and $A^{ll'}_\mu(t,z)
\equiv \tilde{A}^{ll'}_\mu(t,z)-\delta^{ll'}\tilde{A}_\mu(t,z)$, respectively.
For the homogeneous system on the transverse directions, $x$ and $y$, 
$A_t$ and $A_z$ can be taken to be independent of $x$ and $y$ in a gauge, 
so that the non-Abelian part vanishes.
Then, the effective fermion action in Eq.~(\ref{eq:action}) becomes
\begin{equation}
S\simeq {\cal N}\int dtdz 
\bar{\varphi}(t,z)i\tilde{\gamma}^\mu \tilde{D}_\mu\varphi(t,z) ,
\label{eq:action2}
\end{equation}
where $\varphi_l(t,z)$ is rewritten as $\varphi(t,z)$.
This action is nothing but that in 1+1 QED, i.e., the Schwinger model.
The exact solution of the effective action for the fermion is known as
\begin{figure}[tb]
   \centering
   \includegraphics*[width=0.42\linewidth]{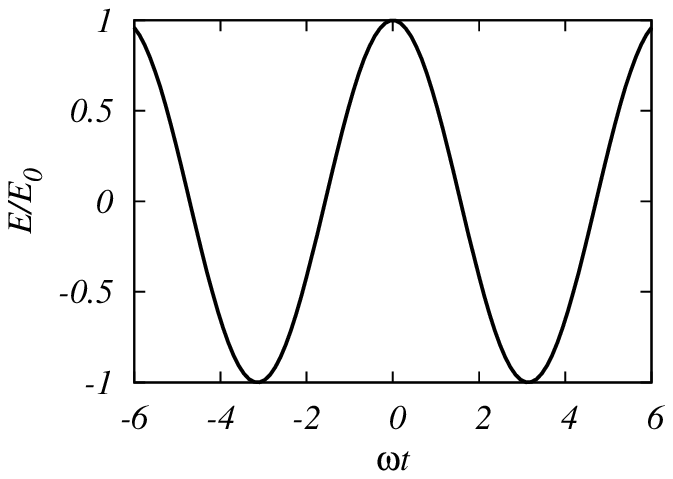}
   \centering
   \includegraphics*[width=0.42\linewidth]{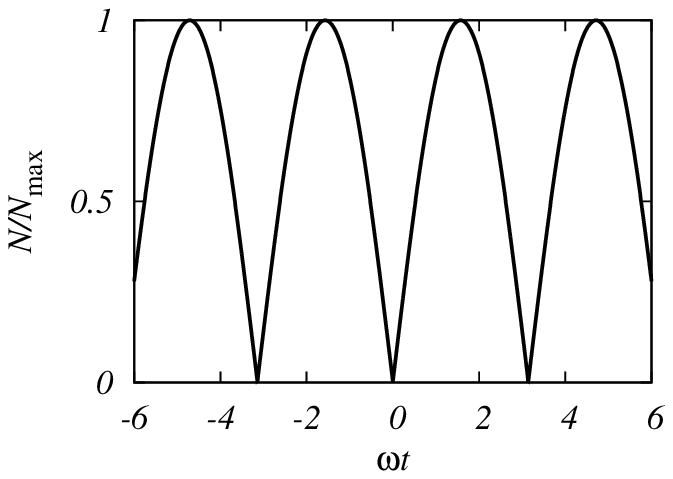}
   \caption{
The electric field $E(t,z)$ (left) and 
the number density $N(t,z)$ of created fermions (right) 
at $z=0$ plotted against time $t$. 
Both values are normalized by that at maximum values.}
   \label{fig:dynamics}
\end{figure}
\begin{equation}
\Gamma(A_\mu)= -\frac{m_\gamma^2V_\perp}{2}\int dtdz \tilde{A}_{\mu}(t,z) 
\left(g_\parallel^{\mu\nu}-\frac{\partial_\parallel^\mu\partial_\parallel^\nu}{\partial_{\parallel}^2}\right)\tilde{A}_\nu(t,z),
\label{eq:effectiveAction}
\end{equation}
where $\parallel$ denotes for longitudinal directions, $t$ and $z$.
$m_\gamma$ is the effective photon mass,
$m_\gamma^2\equiv e^3B/(2\pi^2)$.
Equation~(\ref{eq:effectiveAction}) is manifestly gauge invariant. 
In the Lorenz gauge,  it reduces to the result by \cite{Schwinger3}.
The mass $m_\gamma$ is induced by the axial anomaly, whose effect is called
 ``dynamical Higgs effect.''
From Eq.~(\ref{eq:effectiveAction}), the fermion and axial currents read
\begin{equation}
j^\mu(x)=\frac{\delta\Gamma(A)}{\delta (eA_\mu)}=-\frac{e^{2}B}{2\pi^{2}}\tilde{A}^\mu(t,z), \qquad
j_5^\mu(x)=-\epsilon^{\mu\nu}j_\nu(x),
\end{equation}
where we choose the Lorenz gauge, $\partial_\mu \tilde{A}^\mu=0$.
The divergence of the axial current leads to
the axial anomaly in $1+1$ dimensions 
except for the overall factor $eB/(2\pi)$:
$\partial_\mu j_5^\mu(x)=e^2B\epsilon^{\mu\nu}\partial_{\mu}\tilde{A}_{\nu}(t,z)/(2\pi^2)
=e^2BE/(2\pi^2).$
This relation is nothing but axial anomaly in $3+1$ dimensions.
Since the effective action (\ref{eq:effectiveAction}) 
has a quadratic form in $\tilde{A}_{\mu}$, 
the equation of motion for the photon can be solved. 
For example, the electric field of $z$ direction is
$E=E_0\cos(\omega t-k_zz )$,
with $\omega=(k_z^2+m_\gamma^2)^{1/2}$. The currents satisfy
$ej^\mu=-\epsilon^{\mu\nu}\partial_\nu E$ and
$ej_5^\mu=\partial^\mu E$.
We show in Fig.~\ref{fig:dynamics} 
the electric field $E$ and the particle number density $N$ 
as functions of time $t$, for the spatially homogeneous case, $k_z=0$.
The electric field oscillates with a frequency $\omega$.
In this case, $j^{~t}=0$, but $j^{~t}_5\neq0$.
The number density of created fermions is $|j_5^{~t}|$.
These results agree with the previous works~\cite{Tanji,Iwazaki}. 
Since $m_\gamma^2$ is the only scale in this theory, 
the time scale of the particle decay is $\sim 1/m_\gamma$.

The generalization to non-Abelian theories is straightforward. 
The fermion determinant becomes the Wess-Zumino-Witten action.

\section{Summary and Concluding Remarks}
In this paper, we have studied the vacuum decay 
in homogeneous color electromagnetic fields.
When the fermion is massless, 
the vacuum persistency probability per unit space-time volume becomes zero, 
and hence $w$ diverges. This divergence originates from 
spectral discretization of transverse directions and the lowest Landau level.
With the LLL projection, 
we have analytically calculated the effective action in this situation.
\bibliographystyle{aipproc}   

\end{document}